\documentclass[twocolumn, nofootinbib]{revtex4}
\usepackage{amsmath,amssymb}
\usepackage{graphicx}
\usepackage{psfrag}
\usepackage{color}
\usepackage{epstopdf}

\usepackage{mathrsfs}
\DeclareSymbolFontAlphabet{\mathrsfs}{rsfs}
\DeclareMathAlphabet{\mathcal}{OMS}{cmsy}{m}{n}

\newcommand{\be}{\begin{equation}}
\newcommand{\ee}{\end{equation}}

\begin{document}
\title{A geometric framework for black hole perturbations}

\author{An{\i}l Zengino\u{g}lu}
\affiliation{Theoretical Astrophysics, California Institute of Technology, Pasadena, California, USA}

\begin{abstract}
Black hole perturbation theory is typically studied on time surfaces that extend between the bifurcation sphere and spatial infinity. From a physical point of view, however, it may be favorable to employ time surfaces that extend between the future event horizon and future null infinity. This framework resolves problems regarding the representation of quasinormal mode eigenfunctions and the construction of short-ranged potentials for the perturbation equations in frequency domain.
\end{abstract}

\maketitle

\section{Introduction}
Black hole spacetimes are typically represented in coordinates in which the time hypersurfaces intersect at the bifurcation sphere and at spatial infinity. Because of the simplicity of these coordinates, black hole perturbation theory is typically studied along such time surfaces \cite{Regge:1957rw, Zerilli70, Bardeen:1973xb, Teukolsky:1973ha}. 

\begin{figure}[ht]
  \includegraphics[width=0.37\textwidth]{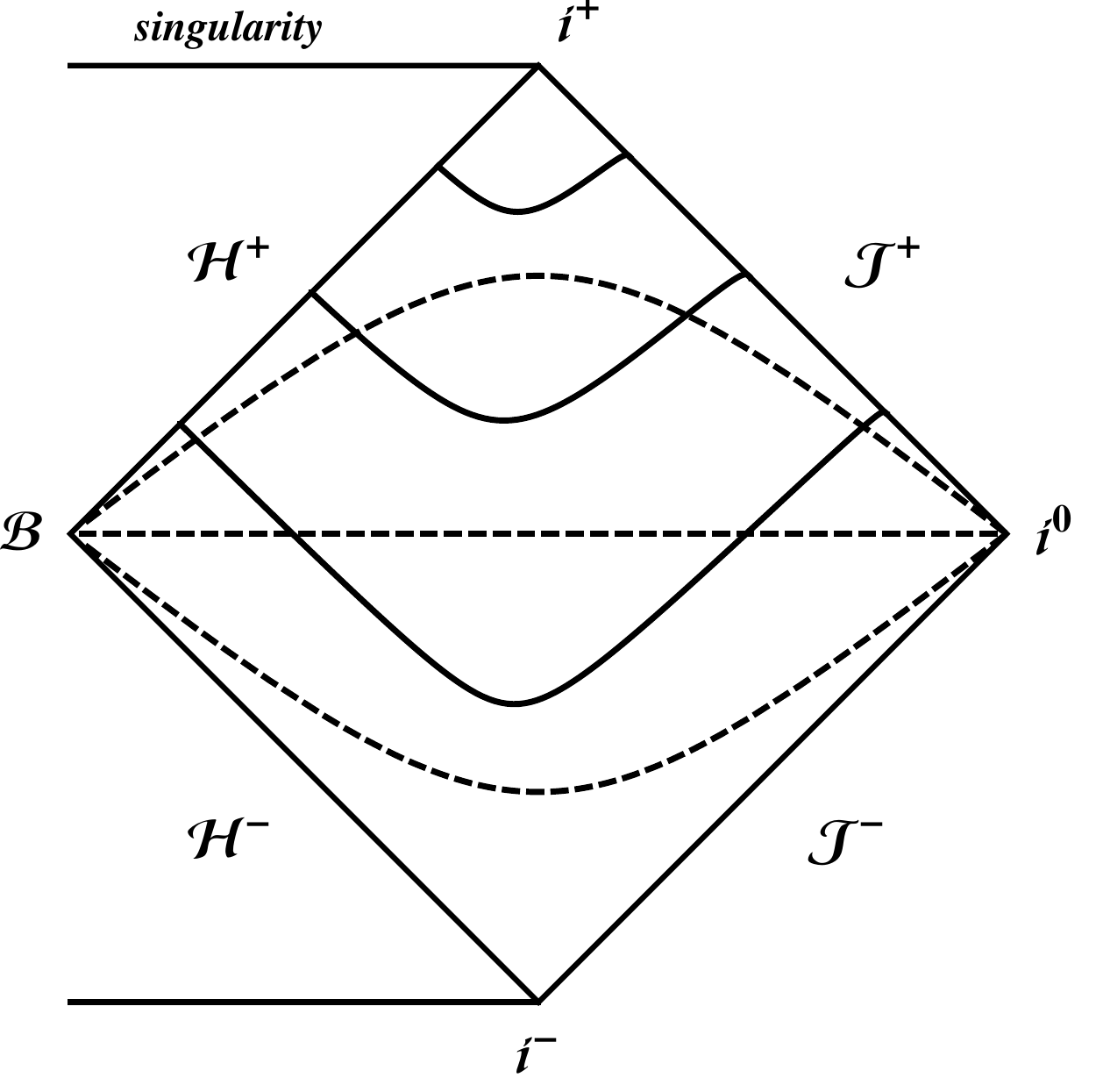}
  \caption{Penrose diagram of the domain of outer communications in Schwarzschild spacetime depicting standard Schwarzschild time surfaces (dashed lines) and future horizon-penetrating, future hyperboloidal  time surfaces (solid lines).\label{fig:conf}}
\end{figure}

In this report, I argue that it is favorable---especially for numerical computations---to employ time surfaces that extend between the future event horizon and future null infinity. These surfaces avoid the pathological behavior at the bifurcation sphere and at spatial infinity. They are called (future) horizon-penetrating and (future) hyperboloidal 
(Fig.~\ref{fig:conf}).

The physical argument is fairly concise. Consider the standard representation of the Schwarzschild metric
 \[ g=-f\,dt^2 + f^{-1}\,dr^2+r^2\,d\Omega^2,  \quad \textrm{with}\quad 
f:=1-\frac{2M}{r}. \] 
Here, $r$ is the areal radius, $d\Omega^2$ is the standard metric on the unit sphere, and $M$ denotes the mass of the black hole. The surfaces of constant time $t$ intersect at the bifurcation sphere, $\mathcal{B}$, and at spatial infinity, $i^0$ (dashed lines in Fig.~\ref{fig:conf}). However, an astrophysical black hole that forms by gravitational collapse does not possess a bifurcation sphere, and observers of gravitational radiation do not have access to spatial infinity. From this point of view, it is appealing to study black hole perturbations on spacelike hypersurfaces extending from the future event horizon, $\mathcal{H}^+$, to future null infinity, $\mathcal{J}^+$ (solid lines in Fig.~\ref{fig:conf}). 

Such hypersurfaces can be constructed using the height function technique \cite{Zenginoglu:2007jw}. Introduce a new time coordinate
\be\label{eq:trafo} \tau(t,x^i) = t - h(x^i),\ee
where the height function, $h$, depends on spatial coordinates only. This transformation leaves the timelike Killing field in the exterior domain invariant. Observers at rest with respect to the black hole have the same representation in the new time coordinate. 
Many useful spacelike foliations of Schwarzschild spacetime, such as Eddington--Finkelstein or Painlev\'e--Gullstrand foliations, have the above form.

It is convenient to employ the tortoise coordinate \mbox{$r_\ast = r + 2M\log[r/(2M)-1]$}. The event horizon in this coordinate is located at $r_\ast\to-\infty$. We choose a height function that depends only on $r_\ast$ and define $H:=dh/dr_\ast$, referred to as the boost function. The Schwarzschild metric becomes
\[ g = f\left(-d\tau^2 - 2H d\tau dr_\ast +\left(1-H^2\right) dr_\ast^2\right) + r^2 d\Omega^2.\]
The boost function for horizon-penetrating, hyperboloidal time surfaces satisfies
\be\label{eq:asymp} |H| \leq 1, \quad \lim_{r_\ast \to \pm\infty} H = \pm 1, \quad  \lim_{r_\ast \to \pm\infty} \frac{dH}{dr_\ast} = 0. \ee
These conditions imply that, asymptotically, the boost of $\tau$-hypersurfaces with respect to $t$-hypersurfaces approaches the speed of light. An example is the boost function of constant mean curvature surfaces \cite{Brill, Malec:2003dq}
\be\label{eq:cmc} H = \frac{J}{\sqrt{J^2+f}}, \quad J := \frac{K r}{3} - \frac{C}{r^2}, \ee
with a positive mean curvature, $K>0$, and $C>8M^3 K/3$ (solid lines in 
Fig.~\ref{fig:conf} with $C=10$ and $K=2$). 

This physically motivated geometric framework has already been applied in numerical studies of gravitational black hole perturbations in the time domain \cite{Zenginoglu:2008uc,Zenginoglu:2009ey,Bernuzzi:2010xj}. In the following, I present a theoretical discussion in the frequency domain. 

\section{Regge-Wheeler-Zerilli equation}
The radial Regge-Wheeler-Zerilli (RWZ) equation describes metric perturbations of Schwarzschild spacetime via multipoles of a Fourier-decomposed master function \cite{Regge:1957rw, Zerilli70}. Dropping the multipole indices, the radial RWZ equation can be written as
\be \label{RWZ} \left(\frac{d^2}{dr_\ast^2} + \omega^2 - U \right) \Psi = 0.  \ee
The potential $U$ depends on whether the perturbations have even or odd parity. Without loss of generality, we consider odd parity perturbations
\[ U = \frac{f}{r^2} \left(\ell(\ell+1) - \frac{6M}{r}\right),\]
where $\ell$ is the angular momentum number. The potential vanishes both at the horizon and at infinity. The asymptotic behavior of the master function in these domains is
\be\label{rwz} \Psi \to C_1 e^{i\omega r_\ast} + C_2 e^{-i\omega r_\ast} \quad \mathrm{as} \quad r_{\ast}\to \pm\infty.\ee
We are interested in solutions that are purely ingoing near the horizon and purely outgoing near infinity, which translates into  $\Psi \to e^{-i\omega r_\ast}$ near the bifurcation sphere and  $\Psi \to e^{i\omega r_\ast}$ near spatial infinity. Solutions with this asymptotic behavior describe exponentially damped oscillations of the black hole, called quasinormal modes (QNMs) \cite{Nollert:1999ji, Kokkotas:1999bd, Berti:2009kk}. 

The frequency $\omega$ has, in general, both a real and an imaginary part. As a consequence, QNM eigenfunctions blow up exponentially near the black hole and near infinity. This behavior is due to the pathological properties of the bifurcation sphere and of spatial infinity: The representation of the physical boundary conditions in these domains is unphysical. 

Switching to horizon-penetrating, hyperboloidal hypersurfaces resolves this problem. The time transformation \eqref{eq:trafo} for the radial RWZ equation in frequency domain is equivalent to the rescaling of the master function
\be\label{rwz_traf} \psi = e^{-i\omega h}\Psi.  \ee
This rescaling, with a specific choice for $h$, is used by Dolan and Ottewill in their expansion method, which has been applied to the calculation of QNMs and Regge poles in the eikonal limit in Schwarzschild and Kerr spacetimes \cite{Dolan:2009nk, Dolan:2010wr}\footnote{I thank Emanuele Berti for pointing out these references.}. It is remarkable that they construct the transformation based on a very different reasoning.

The transformation \eqref{rwz_traf} leads to the following equation
\be\label{trafRW} \left(\frac{d^2}{dr_\ast^2} + 2 i \omega H \frac{d}{dr_\ast} + \omega^2(1-H^2) + i\omega H' - U  \right) \psi = 0.  \ee
The asymptotic form of this equation reads by \eqref{eq:asymp}
\be\label{asympt} \left(\frac{d^2}{dr_\ast^2} \pm 2i\omega \frac{d}{dr_\ast}\right)\psi = 0, \ee
with solutions
\[ \psi \to C_1 e^{\mp 2i\omega r_\ast} + C_2 \quad \mathrm{as} \quad r_{\ast}\to \pm\infty.\]
The physical boundary condition with respect to the new foliation is simply that the solution is of order unity both at the horizon and at infinity, that is, $\psi\to \mathcal{O}(1)$ at both asymptotic ends. As a consequence, the rescaled eigenfunctions are regular in the infinite domain extending from the event horizon to null infinity. This feature extends in a straightforward way to perturbations with even parity and arbitrary spin.

\section{Teukolsky equation}
The radial Teukolsky equation describes curvature perturbations of Kerr spacetimes \cite{Teukolsky:1973ha}. A practical difficulty in numerical computations with the radial Teukolsky equation is its long-ranged potential. Among the various efforts to construct an equivalent equation with a short-ranged potential \cite{PT, TP, Chandrasekhar:1976zz, Detweiler:1977gy}, the most common approach in numerical calculations relies on a generalization of the Chandrasekhar transformation \cite{Chandra} devised by Sasaki and Nakamura \cite{SasakiSS,Sasaki:1981kj}. 

Sasaki and Nakamura reformulated the Teukolsky equation in the RWZ form, first for Schwarzschild spacetime \cite{SasakiSS} and then for Kerr spacetime \cite{Sasaki:1981kj}. Their formalism was restricted to the gravitational case with spin $s=-2$. Hughes later generalized the formalism to arbitrary spins in Kerr spacetime \cite{Hughes:2000pf}. The Sasaki-Nakamura transformation has been applied in various numerical calculations concerning the motion of particles in Schwarzschild and Kerr spacetimes \cite{Sakamura, Hughes:1999bq, Finn:2000sy, Cardoso:2002yj}.

The framework presented in the introduction leads to a short-ranged potential with a simple transformation that has a clear physical and geometric interpretation. We discuss its application to the Teukolsky equation first in Schwarzschild spacetime, and then in Kerr spacetime, for arbitrary spins.

\subsection{Schwarzschild spacetime}

The radial Bardeen-Press-Teukolsky (BPT) equation in Schwarzschild spacetime for perturbations of spin weight $s$ reads \cite{Bardeen:1973xb,Teukolsky:1973ha}
\be\label{BPT} \left(\frac{d^2}{dr_\ast^2} + \frac{2\left( (r-M)(1+s)-M\right)}{r^2} \frac{d}{dr_\ast} + \omega^2 - U\right) \Psi = 0, \ee
with the potential
\[ U = -\frac{2i\omega s}{r^2}(r-3M) + \frac{f}{r^2}(\ell(\ell+1)-s(s+1)). \]
The problematic term in \eqref{BPT} is the first term of the potential that does not vanish at the horizon and falls off as $r^{-1}$ near infinity. To construct a short-ranged potential, one that vanishes at the horizon and falls off at least as $r^{-2}$, we rescale the master function according to its asymptotic behavior. The perturbations fall off as $r^{-(2s+1)}$ near infinity, and as $f^{-s}$ near the black hole \cite{Teukolsky:1973ha}. A further rescaling transforms the equation to a horizon-penetrating, hyperboloidal foliation as in \eqref{rwz_traf}. We set
\be \label{rescbpt} \psi = r^{-(2s+1)}f^{-s} e^{-i \omega h} \Psi. \ee
The modified BPT equation reads
\begin{eqnarray} \label{BPTHyp}
\frac{d^2\psi}{dr_\ast^2} + \left(-\frac{2s}{r^2} (r-M)+2i\omega H\right) \frac{d\psi}{dr_\ast}  \nonumber \\
+\left(\omega^2(1-H^2) + i\omega H' - \tilde{U}\right)\psi = 0,
\end{eqnarray}
with the potential
\begin{eqnarray}\label{potHyp} 
\tilde{U} &=& - \frac{2i s \omega}{r^2}\left(r f (1-H) - M(1+H) \right)  \nonumber \\ 
&&  + \frac{f}{r^2}\left(\ell(\ell+1)-s(s+1) +\frac{2M}{r}(s+1)\right).
\end{eqnarray}
The asymptotic behavior of the boost function given in \eqref{eq:asymp} implies that this potential falls off as $f$ near the horizon and as $r^{-2}$ near null infinity. Therefore, the modified potential is short-ranged. Note that all lowest order terms in \eqref{BPTHyp} vanish at the asymptotic ends. 

\subsection{Kerr spacetime}
The radial Teukolsky equation in Boyer-Lindquist coordinates reads
\[ \Delta^{-s} \frac{d}{dr} \left(\Delta^{s+1} \frac{d\Psi}{dr}\right) + U \,\Psi = 0, \]
where $\Delta := r^2+a^2-2M r$, and
\[ U = \frac{K^2-2 i s(r-M) K}{\Delta} + 4 i s \omega r - \lambda,  \]
with $K:= (r^2+a^2)\omega-m a$, and
\[ \lambda: = \mathcal{E}_{\ell m} - 2 a m \omega + a^2\omega^2 - s(s+1).\]
Here, $\mathcal{E}_{\ell m}$ is the eigenvalue of the spherical harmonic \cite{Hughes:1999bq} (it becomes  $\mathcal{E}_{\ell m} = \ell(\ell+1)$ in the Schwarzschild limit).

We write the Teukolsky equation in the tortoise coordinate to make the connection to the previous sections. The tortoise coordinate in Kerr spacetime is defined via
\[ dr_* = \frac{r^2+a^2}{\Delta} \,dr. \]
The Teukolsky equation in the tortoise coordinate reads
\[ \frac{(r^2+a^2)^2}{r^4} \frac{d^2\Psi}{dr_*^2} +\frac{2 G}{r^4} \,\frac{d\Psi}{dr_*}  + \frac{\Delta}{r^4} U \Psi = 0, \]
where
\[ G = M(a^2-r^2)+(r-M) (r^2+a^2)(s+1).\]

A geometric transformation that leads to a short-ranged potential can be given as
\be \label{resctk} \psi = r^{-1}\Delta^{-s} e^{i m \tilde{\phi}} e^{-i \omega h} \Psi.\ee
This transformation is very similar to \eqref{rescbpt} with one essential difference: We transform also the angular coordinate $\phi$. The Boyer-Lindquist angular coordinate $\phi$ leads to pathologies near the horizon \cite{Krivan:1996da}. To cure this behavior, we introduce the angular coordinate $\tilde{\phi}$ defined by
\be\label{angular} d\tilde{\phi} =  d\phi + \frac{a}{r^2+a^2} dr_*, \ee
This azimuthal transformation is used already by Teukolsky in his seminal paper on rotating black hole perturbations to derive the boundary conditions near the horizon \cite{Teukolsky:1973ha}. It is commonly used in numerical computations in time domain codes \cite{Krivan:1996da,Sundararajan:2007jg}.

The Teukolsky equation for the transformed eigenfunction $\psi$ from \eqref{resctk} reads
\[ \frac{(r^2+a^2)^2}{r^4} \frac{d^2\psi}{dr_*^2} - \frac{2 \widetilde{G} }{r^5}\,\frac{d\psi}{dr_*}  + \frac{\tilde{U}}{r^6} \psi = 0, \]
with
\[ \widetilde{G} = a^2 \Delta + (r^2+a^2) \left(r s(r-M)- i r ((r^2+a^2)\omega H + m a)\right),\]
and
\begin{eqnarray*}
\tilde{U} &=& 2 i s\omega r^2 \left(  r \Delta (1-H) - M(r^2-a^2)(1+H) \right) \\ 
&& - 2 i a r \Delta(m+a \omega H) + \Delta \left(2a^2 - r^2\lambda - 2Mr(s+1) \right) \\ 
&& - 2 m a \omega r^2(r^2+a^2)(1+H)\\
&& + r^2 (r^2+a^2)^2 \left( \omega^2(1-H^2) + i \omega H' \right).
\end{eqnarray*}
The lowest order term $\tilde{U}/r^6$ vanishes as $\Delta$ at the horizon and falls off as $r^{-2}$ near null infinity due to the asymptotic behavior of the boost function \eqref{eq:asymp}. So the rescaled Teukolsky equation in horizon-penetrating, hyperboloidal coordinates with the angular coordinate $\tilde{\phi}$ of \eqref{angular} has a short-ranged potential.

\section{Discussion}
The framework described in this report has advantages over standard methods in black hole perturbation theory. The QNM eigenfunctions have a finite representation in an infinite domain, and the radial Teukolsky equation becomes short-ranged with the transformation \eqref{resctk}, which is simpler than the Sasaki-Nakamura transformation, and has a clear physical and geometric interpretation.
\pagebreak

This method may lead to efficient numerical codes in the frequency domain, especially in combination with compactification of the radial coordinate. Compactification would remove the cumbersome asymptotic expansions and the studies of inner and outer boundary effects in the numerical integration of ordinary differential equations appearing in frequency domain computations. This technique should be studied in future research on astrophysically motivated examples, such as calculations of extreme mass ratio inspirals using the Teukolsky formalism, or the self-force approach.

The geometric origin of this framework is a clear advantage. It implies that one does not need to apply the transformation \eqref{resctk} back and forth in solving the Teukolsky equation numerically, as it is done with the Sasaki-Nakamura transformation \cite{Hughes:1999bq, Finn:2000sy}. One can directly solve the transformed equation, and extract the physical quantities, such as energy fluxes, at null infinity from the rescaled master function without going back to the Teukolsky equation in Boyer-Lindquist coordinates.

The new framework can also be useful in analytic studies. Dolan's expansion method as applied to Kerr spacetime is restricted to equatorial and polar modes \cite{Dolan:2010wr}. The geometric framework presented in this report indicates that the angular transformation \eqref{angular} may lead to an extension of the expansion method to general modes in Kerr spacetime. It would also be interesting to study how this framework relates to Green function based regularization techniques of divergent integrals for the Teukolsky equation with sources \cite{DetSz, Tashiro:1981ae, Poisson:1996ya}.


\section*{Acknowledgments}
I thank Emanuele Berti, Piotr Bizo\'n, Yanbei Chen, Sam Dolan, and Tanja Hinderer for discussions. This research was supported by the NSF grant PHY-0601459 and by a Sherman Fairchild Foundation grant to Caltech.

\end{document}